# AI Overviews in Academic Search: Evaluating AI-generated Summaries of Search Results in a Domain-specific Search Engine

**Schott, Kevin** GESIS – Leibniz Institute for the Social Sciences, Germany | kevin.schott@gesis.org

**Silva, Kanishka** GESIS – Leibniz Institute for the Social Sciences, Germany | kanishka.silva@gesis.org

**Frommholz, Ingo** Modul University Vienna, Austria | ifrommholz@acm.org

**Mayr, Philipp** GESIS – Leibniz Institute for the Social Sciences, Germany | philipp.mayr@gesis.org

**Kern, Dagmar** GESIS – Leibniz Institute for the Social Sciences, Germany | dagmar.kern@gesis.org

**Hienert, Daniel** GESIS – Leibniz Institute for the Social Sciences, Germany | daniel.hienert@gesis.org

## ABSTRACT

Evaluating search engine results pages (SERPs) to assess result relevance is a demanding step in academic search. In a formative mixed-methods design study, we examine AI-generated SERP-level summaries as a support feature in an academic search engine for social science information. First, we manually evaluated summaries of the top five results for 10 queries using two general-purpose models, one commercial and one open, deriving an exploratory six-category error taxonomy and five safeguards for scholarly deployment. We then conducted a within-subjects user study ($n$ = 30) comparing interfaces with and without AI summaries. Confirmatory analyses showed consistent but non-significant trends favoring AI summaries for subjective workload, perceived usefulness, satisfaction, and decision-making confidence. Exploratory analyses suggested lower mental demand, with frustration also tending to be lower. Behaviorally, participants rarely expanded the summaries and descriptively made slightly fewer result clicks and query reformulations when summaries were available. Drawing on Information Foraging Theory and participant feedback, we suggest that AI summaries may concentrate SERP-level information scent to support early triage. Overall, the findings indicate that SERP-level AI summaries are a context- and user-dependent aid rather than a universal improvement, while contributing an error taxonomy, safeguard-aware deployment guidance, and concrete design implications for scholarly search.

## INTRODUCTION

Researchers today face an ever-expanding volume of academic publications and research data (Bawden & Robinson, 2009; Extance, 2018; Hołyst et al., 2024; Tenopir et al., 2003). In academic search, this abundance creates the challenge of rapidly triaging and assessing relevance under uncertainty using limited cues on the search engine results page (SERP), such as titles, abstracts, and other metadata (Marshall & Shipman, 1997). Prior work on academic information seeking highlights that early-stage searching is often characterized by uncertainty and evolving problem understanding, and that users can benefit from support that helps them orient and refine their focus (Kuhlthau, 1991). In high-volume contexts, this triage burden can also contribute to feelings of overwhelm and information overload (Bawden & Robinson, 2009; Belabbes et al., 2022; Roetzel, 2019).

AI-powered tools offer the possibility to synthesize cues that can help users assess and organize result sets more efficiently (Extance, 2018; Grainger et al., 2025). In this work, we examine a specific interface intervention: inserting a single, AI-generated abstractive summary of the five top-ranked results at the top of the SERP in an academic search engine for social science information. Our research question is whether an off-the-shelf, general-purpose LLM (without domain-specific fine-tuning) can generate useful multi-document summaries that support researchers during data discovery while preserving source-oriented scholarly workflows.

Prior work on extractive single-document and single-result summaries shows benefits for efficiency and relevance assessment (Alexander et al., 2000; Tombros & Sanderson, 1998; White et al., 2003). Recent advances in LLMs enable higher-quality abstractive synthesis across multiple sources (Godbole et al., 2025; H. Zhang et al., 2025; Y. Zhang et al., 2025). At the same time, studies of generative AI in general web search highlight a trade-off: efficiency gains can coincide with shallower engagement and fewer visits to primary sources (Kaiser et al., 2025; Lee et al., 2025; Yang et al., 2025). However, our setting differs: users of academic search engines explicitly seek source documents and must critically appraise and verify them rather than relying on one-shot answers.

While extractive single-result summaries have been studied for decades, the paradigm of a single, AI-generated summative overview spanning multiple results remains under-investigated in specialized scholarly contexts. It is unclear how SERP-level summaries affect triage behavior and whether they can support users. We address this gap through a formative mixed-methods design study combining two investigations: (1) a manual quality assessment comparing two general-purpose models and deriving an exploratory error taxonomy with corresponding safeguards, and (2) a controlled user study examining subjective, behavioral, and qualitative effects of SERP-level AI summaries, combining confirmatory hypotheses with exploratory analyses given the formative scope of the work. Drawing on Information Foraging Theory, we interpret our findings through the idea that such summaries may act as a "scent concentrator," compressing cues from multiple top-ranked results into a compact, source-linked representation that supports early-stage triage by reducing the cognitive cost of entering the information patch.

Overall, our findings point to user- and context-dependent benefits rather than a universal advantage. This work contributes empirical evidence, design implications, and safeguard-aware guidance for integrating AI summaries into scholarly search interfaces.

## RELATED WORK

### Extractive, Abstractive, and LLM-based Summarization

Search engine result lists commonly include short, query-focused summaries (snippets) of individual search results to help users evaluate the relevance of documents (Schütze et al., 2008). This practice stems from Automatic Text Summarization (ATS), a field dedicated to distilling information from one or more sources into concise, task-specific summaries, with query-focused summarization being one particular approach (Abdul Salam et al., 2025; Alanzi & Alballaa, 2023; El-Kassas et al., 2021; H. Zhang et al., 2025). Historically, this field has been dominated by extractive summarization, which selects key sentences directly from a document. While this approach reduces the risk of factual errors by copying source sentences, it can lack semantic cohesion. In contrast, abstractive summarization generates novel sentences to represent the source content, allowing for greater conceptual coherence, but introducing the risk of factual inaccuracies or "hallucinations" (El-Kassas et al., 2021; H. Zhang et al., 2025; Y. Zhang et al., 2025).

The advantages of Large Language Models (LLMs) have transformed ATS, inducing a shift toward high-quality abstractive summarization, often via zero-/few-shot prompting, without the need for large, domain-specific training datasets (H. Zhang et al., 2025; Y. Zhang et al., 2025). A key capability of long-context LLMs is their enhanced performance in multi-document summarization (MDS), which allows for the synthesis of information from several sources into a single summative overview that can include citations (Godbole et al., 2025; Grainger et al., 2025) (i.e., traceable summarization (Chu et al., 2025)). Modern web search engines such as Google or Bing, as well as academic search tools such as ScopusAI (https://elsevier.libguides.com/Scopus/ScopusAI, last accessed: April 2, 2026), present AI-generated overviews that synthesize across multiple sources and link to citations, intended to simplify users' assessment of the content. Thereby, retrieval-augmented generation (RAG) is utilized as a central grounding approach (Chen et al., 2024; Grainger et al., 2025; Withorn, 2025).

However, despite these advances, challenges such as hallucinations, biases, and incorrect instruction-following of the LLM remain (Chen et al., 2024; Chu et al., 2025; Li et al., 2024; Y. Zhang et al., 2025). These may be particularly pronounced in specialized domains such as academic search, where documents are often dense and contain complex jargon. To investigate the efficacy of summaries in this high-stakes context, our study utilizes RAG and the MDS capability of LLMs to generate a unified, abstractive summary of the top five search results in a search engine for social science information.

### Summarization of Scientific Documents

As researchers grapple with an ever-expanding volume of publications, AI-powered tools provide the opportunity of synthesizing this vast information to make it more easily digestible (Extance, 2018). Automatic summarization of scientific documents is a particularly promising area that presents unique challenges due to the texts' inherent length, structure, and domain-specific language (Ibrahim Altmami & El Bachir Menai, 2022; Zaman et al., 2024). To address this, specialized approaches have been developed. *Faceted summarization*, for instance, generates structured outputs by summarizing distinct sections such as a paper's purpose, methods, and findings (Meng et al., 2021). The demand for rapid content assessment has also driven research into *extreme summarization*, which produces single-sentence summaries of a paper's core contribution, facilitated by novel datasets (Cachola et al., 2020; Mao et al., 2022). Finally, some work has focused on combining summarization with *simplification*, aiming to make scientific findings more accessible to non-experts by generating summaries that are not only shorter but also lexically simpler (Zaman et al., 2024).

In contrast to these specialized approaches, our study investigates a more generalized application reflective of current AI capabilities. We explore whether a general-purpose LLM can produce useful multi-document summaries when prompted with the metadata of the five top-ranked search results. This allows us to assess a non-specialized, "off-the-shelf" approach, examining its effectiveness in a realistic academic search scenario without relying on domain-specific fine-tuning.

### Effects of Query-biased Extractive Summaries for Single Search Results

A foundational body of research has established that providing users with high-quality extractive summaries alongside single search results improves both the efficiency and effectiveness of information seeking. Studies have shown that compared to generic snippets (e.g., the first few sentences of a document), query-biased summaries can significantly increase search precision and recall (Tombros & Sanderson, 1998), reduce the number of queries issued and the time required to complete tasks (White et al., 2003), and decrease the need for users to open full documents to assess relevance (Alexander et al., 2000; Tombros & Sanderson, 1998). This can lead to higher user satisfaction with their search results when provided with query-biased summaries (White et al., 2003). Compared to generic snippets or a title-only approach, users also perceive query-biased summaries as more useful for assessing the relevance of results (White et al., 2003) and for answering a search task's questions (Alexander et al., 2000). The effectiveness of these summaries is also tied to their interpretability, as search results whose summaries are more transparent (regarding why a result was retrieved) and assessable (easy to tell if a result will be useful) are more likely to be clicked (Mi & Jiang, 2019).

This body of work provides a robust understanding of user interaction with traditional, extractive summaries for single search results. However, it does not address the paradigm enabled by modern LLMs: the generation of a single, abstractive summary synthesized from multiple top-ranked documents. Our research aims to bridge this gap. By integrating an AI-generated summary module based on the user's query and top five search results, our study investigates how the presented findings apply to this new form of multi-document search result presentation in a specialized academic context.

### Influence of Generative AI on Information Seeking Behavior

Recent research on Generative AI (GenAI) in the context of information seeking reveals a trade-off between efficiency and the depth of user engagement. Both conversational chatbots (Kaiser et al., 2025) and non-conversational LLM-based search tools (Spatharioti et al., 2025) have been shown to decrease completion time for search tasks and lead to enhanced perceived search experiences (Spatharioti et al., 2025). Qualitative feedback also indicates that users value saving time and energy with conversational AI assistance (Yang et al., 2025). However, this perceived benefit may correspond with reduced deeper engagement: in a survey of knowledge workers, higher confidence in GenAI's ability on the task predicted lower task-specific engagement in critical thinking (Lee et al., 2025). While the increased speed can lead to higher success rates in search tasks (Kaiser et al., 2025), accuracy may decrease significantly when the LLM outputs incorrect facts on deliberately challenging items (Spatharioti et al., 2025).

These tools often foster a "closed-loop" interaction where users visit significantly fewer primary sources, treating the AI as a one-stop information source (Kaiser et al., 2025; Yang et al., 2025). A qualitative study with 21 expert users observed substantially fewer source interactions with an LLM-based answer engine compared to a traditional search engine. The authors provide design recommendations and an evaluation benchmark that aim to mitigate limitations such as unbalanced viewpoints and citation misattributions (Narayanan Venkit et al., 2025). This restricted engagement with primary sources leads to concerns about overreliance, as users frequently do not verify the AI's fluent answers, resulting in misplaced confidence and poor task accuracy, especially when the model hallucinates (Mayerhofer et al., 2025; Spatharioti et al., 2025).

Complementing this behavioral research, recent work has begun examining visual attention patterns. Rolon-Merette et al. (Rolon-Merette et al., 2025) compared Google SERPs with "AI Overviews" to standard SERPs using eye-tracking. Although the overviews drew greater visual attention and were perceived as useful, they did not affect click-through rates or physiologically measured cognitive load in this study. Building on such approaches, Al Lawati (Al Lawati, 2025) proposes eye-tracking studies to investigate how users navigate when AI-generated content appears above the traditional "ten blue links," examining scrolling behaviors and whether interaction patterns differ from established literature on search interface scanning.

Collectively, these works demonstrate that while GenAI offers significant efficiency gains, it introduces risks of shallower engagement and over-reliance. Most of this research, however, has investigated conversational agents or direct-answer formats in general web search contexts. It is less clear how these dynamics will manifest in a specialized academic domain like the social sciences where researchers search for source documents and where verification and critical evaluation are paramount. Our study addresses this gap by investigating a multi-document summary feature embedded within a traditional search interface for a scholarly audience.

## RESEARCH QUESTIONS

While the studies reviewed above demonstrate GenAI's potential for efficiency in general search, they leave open questions about its application in academic contexts with multi-document summaries where the objective is to identify relevant information sources for further investigation rather than to find a direct answer to a query. To address this, we pose the following research questions:

- How faithfully do AI-generated SERP-level summaries represent SERP metadata in academic search, and what errors occur most frequently?
- How do AI-generated SERP-level summaries affect early-stage information triage by social scientists in an academic search engine?

To investigate these questions, we evaluate the quality of AI-generated summaries in isolation, before integrating them into an academic search engine for social science information to conduct a user study. First, we describe the search engine and the design of the AI-generated summary in greater detail in the following section.

## ACADEMIC SEARCH ENGINE & AI-GENERATED SUMMARY

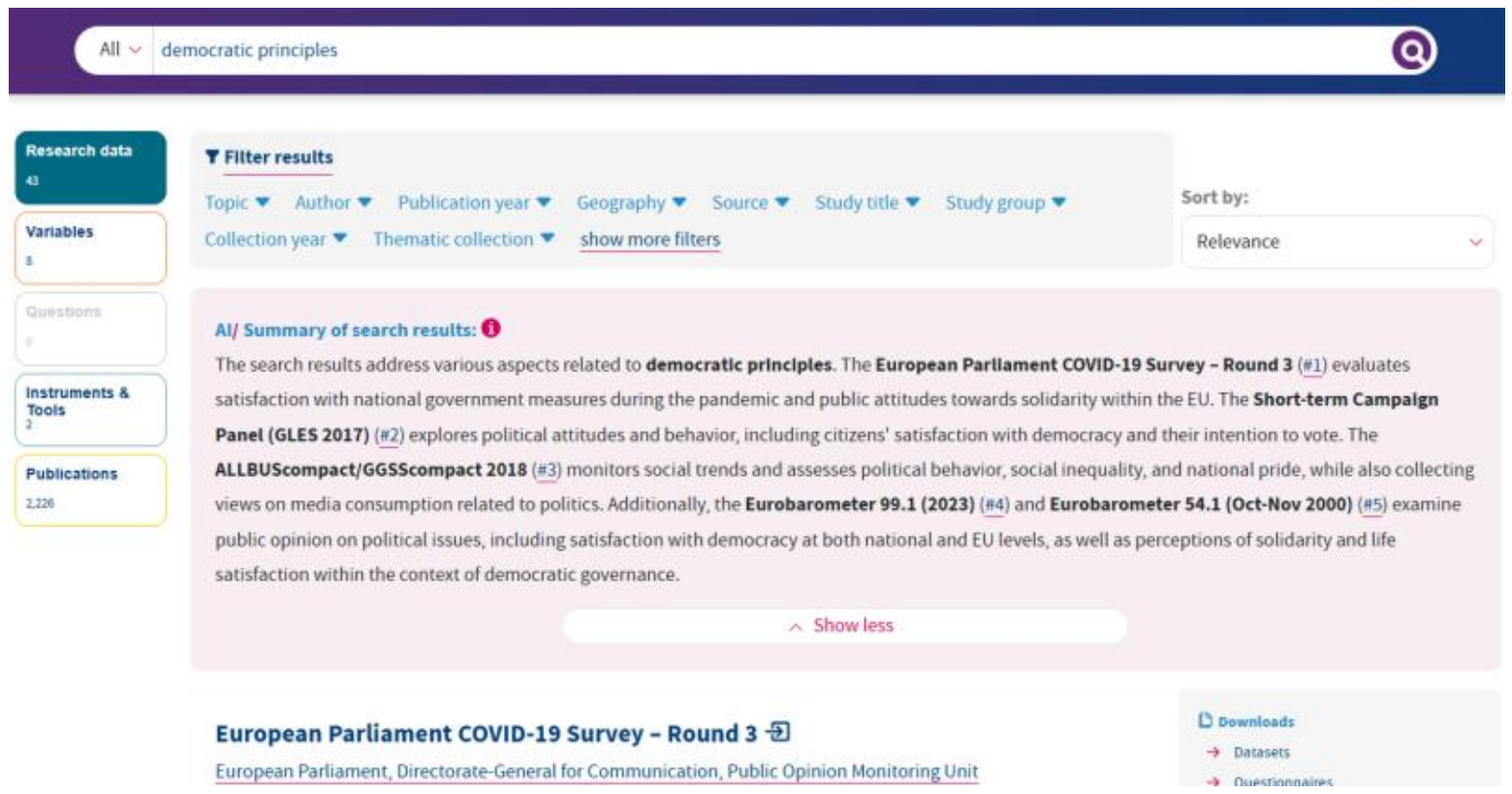

**Figure 1. Screenshot of the *AISummary* Condition with Expanded Summary for an Exemplary Query**

We conducted our study with an academic search engine for social science information (see Figure 1). The system uses keyword-based retrieval and its index integrates research datasets (~7,500 quantitative studies from 1945–2025) and ~250,000 scholarly publications, among other information. Records are interlinked (e.g., publications with data citations to underlying datasets), faceted filters support refined retrieval, individual search results can be bookmarked, and depending on different licensing models, data and materials can be downloaded. For each search result, the search engine also provides an abstract preview on the SERP.

Importantly, the use case of an academic search engine significantly differs from that of a non-academic or general web search engine (Gusenbauer, 2019; Kacprzak et al., 2019; Ortega, 2014). Thus, while web searchers often seek a direct answer to a question, users of an academic search system aim to find relevant source documents (i.e., datasets or publications) to help answer a research question. Accordingly, our design aimed to provide a high-level query-focused summative overview of the five top-ranked search results rather than a direct answer to the user's query.

To generate abstractive AI summaries for academic search results, we instructed the LLM to act as a research assistant and to write a single coherent paragraph of no more than five simple sentences synthesizing all results listed in the prompt (the prompt we used to generate the summaries is available at https://osf.io/gjhky/overview?view_only=e82c658aa3a04d39b0a931ab911b8509). The prompt further required the model to mention each item and cite its identifier (displayed on the interface as a clickable number matching its position in the list, e.g., '#1' or '#2'), follow formatting rules that highlight query terms and each result's unique contribution, and group together items whose content substantially overlaps. Finally, it specified that the model must avoid deductions or interpretations and rely only on information contained in the provided results, without including external knowledge. Once generated, a summary for a particular query was cached for all subsequent study sessions. Figure 1 shows an example of the generated summary for the user query "democratic principles."

## AI SUMMARY QUALITY EVALUATION (RQ1)

To evaluate summary quality for display in an academic search engine, we conducted a manual metadata-grounded accuracy check, addressing RQ1. We compared a commercial model, GPT-4o mini, to an open, locally hosted model, Llama 4 Scout (67GB, 109B parameters, Q4_K_M). These two models were chosen to represent two common "off-the-shelf" deployment options without domain-specific fine-tuning. We used the following ten sample keyword queries, sourced from social scientists (Papenmeier et al., 2021), to generate summaries: "psychosocial stress", "social background", "work equipment employees", "dimensions of social identity", "marriage transfers", "politically motivated crime", "educational attainment refugees", "democratic principles", "discrimination muslims", and "sampling mobile phone survey." Following a qualitative evaluation framework (available at https://osf.io/gjhky/overview?view_only=e82c658aa3a04d39b0a931ab911b8509) iteratively developed on separate exemplary queries, three independent raters (the first, fifth, and sixth authors) evaluated each summary's faithfulness to the SERP metadata. After familiarizing themselves with the framework, raters assigned "Correct," "Partially

Correct," or "Incorrect" labels using its hierarchical rules for metadata representation and unsupported claims. For each detected inaccuracy, the raters provided a comment describing the type of error (e.g., "wrong reference"). We adopted a consensus-based approach in which the three raters discussed and resolved all disagreements to reach a shared final label for each summary, following best practices for interpretive coding tasks (McDonald et al., 2019).

### GPT-4o mini Evaluation

To characterize inaccuracies in the "Partially Correct" and "Incorrect" summaries ($N = 5$), we conducted a qualitative thematic analysis (Braun & Clarke, 2006) in which one researcher (first author of this paper) inductively coded the three raters' comments. The following three error themes emerged, with representative rater comments and frequencies ($n$) integrated: *Inaccurate or Unsupported Content Descriptions*–summaries introduce details not present in, or misrepresent, the source metadata ("'Marriage transfers' is not explicitly mentioned in the metadata, only 'marriage'. Therefore labeled as inaccurate."; $n = 3$); *Unsubstantiated Guesses or Speculations*–the model infers connections to the user's query without evidence in the metadata ("'which may provide contextual insights…' is only guessed–triggered by prompting the model to make a relationship to the user query."; $n = 3$); and *Omission of Search Results*–one or more provided results are left out of the summary ("Result 5 (Social Security & Fairness): Missing, therefore labeled as inaccurate."; $n = 2$). The evaluation of the GPT-4o mini-generated summaries resulted in 5 of the 10 summaries being labeled as "Correct," 4 as "Partially Correct," and 1 as "Incorrect."

### Llama 4 Scout Evaluation

We generated summaries using an open model, Llama 4 Scout, for the same 10 social science queries, applied the same evaluation framework, and compared the resulting quality assessments between the two models. The consensus labels for Llama 4 Scout were 2 "Correct" and 8 "Partially Correct," with no "Incorrect," indicating notably lower overall quality compared to GPT-4o mini. A similar qualitative thematic analysis was conducted on the raters' comments for the "Partially Correct" summaries ($N = 8$). One overlapping theme between the two models emerged–*Inaccurate or Unsupported Content Descriptions* ("That's not in the metadata: 'which highlights the significance of social origin…'."; $n = 3$)–along with three new themes introduced with representative rater comments and frequencies ($n$): *Incorrect Study Identifiers* (wrong or invented study numbers in citations; "Incorrect mention of ZA7730 (Wrong study number)."; $n = 4$), *Formatting and Language Issues* (e.g., typos or unnecessary subheadings; "Unnecessary list of topics at the end of the summary."; $n = 2$), and *Inappropriate Grouping of Studies* (illogical summarization of unrelated studies together; "Doesn't make sense to me to summarize these studies together."; $n = 2$). Notably, the themes *Unsubstantiated Guesses or Speculations* and *Omission of Search Results* observed for GPT-4o mini did not appear in Llama 4 Scout, suggesting model-specific error patterns.

Because GPT-4o mini produced fewer practically serious errors, we selected it for the experimental condition in the user study.

## USER STUDY ON AI SUMMARY (RQ2)

To address RQ2, we conducted a controlled remote user study employing a within-subjects experimental design. Data collection took place from 1–22 September 2025. We invited participants to single online study sessions on Microsoft Teams.

### Interface Variants

We evaluated two variants of the academic search engine for social science information: *NoSummary* and *AISummary*. Both variants featured a persistent query bar, a vertical list of search results, and facets to switch between resource types such as publications (i.e., scholarly literature) and research data (i.e., datasets) (see Figure 1). Each participant interacted with both variants, and the order of interaction was counterbalanced:

- **NoSummary (Baseline):** Participants used the standard interface of the search engine as described above.
- **AISummary (Experimental):** This condition augmented the baseline interface with an expandable pane displaying a query-focused abstractive summary of the five top-ranked search results. In the default collapsed state, users could see four lines (i.e., about half) of the full summary. The summaries were generated in real time by OpenAI's GPT-4o mini model (time of use: 1–22 September 2025) based on the results' metadata (i.e., title, abstract, and topics/keywords).

To increase ecological validity by covering both familiar and unfamiliar academic search contexts, each participant completed both a self-chosen **familiar topic** task and a randomly assigned **predefined topic** task, randomly paired (with equal distribution) with the two interface conditions:

- **Familiar Topic:** *"Please use the search system to find research data and/or publications on a topic you are already familiar with (e.g., your main research area). Bookmark relevant results to create a watchlist."*

- **Predefined Topic:** *"Please use the search system to find research data and/or publications on this topic. Bookmark relevant results to create a watchlist."*

We randomly selected the predefined topic from one of the following three options with equal distribution: (i) Citizens' attitudes toward democracy in various European countries, (ii) Experiences of discrimination among EU immigrants, or (iii) Social factors influencing perceptions of stress in various European countries.

### Procedure and Tasks

The study moderator (first author of this paper) guided the participants through the process via Microsoft Teams. We recorded the audio of the session and the participant's screen via Teams' screen sharing feature, and automatically generated transcripts of the sessions. To ensure anonymity, participants were instructed to change their display names and to turn off their cameras for the study.

The study started with an online questionnaire hosted on SoSci Survey (https://www.soscisurvey.de/en/index). Participants first provided informed consent before receiving a brief introduction to the search system. Subsequently, they completed both tasks using both interfaces in their assigned order. For each search task, participants received a link in the questionnaire to open the respective interface condition in a new tab. Each participant was instructed to use the search system's bookmark functionality to add relevant search results to a personal watchlist. Clicks to open search results, bookmarks, and query (re)formulations were captured via automatic system logging with timestamps and participant IDs for each task. The study moderator ended each search session after 10 minutes. This 10-minute boundary was chosen deliberately to isolate and observe the early-stage triage phase.

After completing both search tasks, participants answered a single-choice interface-preference item (*AISummary* / *NoSummary* / no preference) and then participated in a brief semi-structured interview. Subsequently, the participants answered demographic questions (age, gender, education level, job title for employees / major field of study for students) and questions about their familiarity with academic search engines and with the tested search system. Participants were instructed to disable screen sharing while completing the questionnaire. The study was approved by our institute's ethics committee.

### Measurements

#### *Self-reported Variables for both Conditions*

After each task, we collected: (1) ***Subjective workload*** using the Raw NASA–TLX (Hart & Staveland, 1988) (six subscales on 21-point scales mapping to 0–100; composite = mean of subscales), (2) ***perceived usefulness*** via Davis' six-item scale (Davis, 1989) (7-point Likert, 1="Extremely unlikely" to 7="Extremely likely"), (3) ***satisfaction*** (single item, 7-point Likert, 1="Strongly disagree" to 7="Strongly agree": "I am satisfied with my search experience"), and (4) ***decision-making confidence*** inspired by (Maxwell et al., 2017) (single item, 7-point Likert, 1="Strongly disagree" to 7="Strongly agree": "I felt confident in deciding which search results to explore further").

#### *Ratings of the AI-generated Summary*

For the *AISummary* condition, we displayed a blurred summary example and measured five aspects on 7-point Likert scales (1="Strongly disagree" to 7="Strongly agree", with an "I don't know" option to account for participants who did not notice and/or read the summary): (1) ***Relevance assessability***, adapted from (Maxwell et al., 2017; Mi & Jiang, 2019; White et al., 2003) ("By looking at the summaries, I could tell whether the search results were relevant without opening the link"), (2) ***appropriateness***, following (Maxwell et al., 2017) ("In general, the summaries were appropriate in size and length"), (3) ***understandability***, following (Maxwell et al., 2017) ("...understandable"), (4) ***perceived usefulness*** ("...useful"), inspired by (Alexander et al., 2000; Sweeney & Crestani, 2006; White et al., 2003), and (5) ***perceived reliability*** ("...reliable"), inspired by (Gulati et al., 2019; Madsen & Gregor, 2000; Sullivan & Weger, 2025).

#### *Interface Preference*

After completing both search tasks, participants answered: "Which of the two interface variants do you prefer?" with options (a) "the variant displaying AI-generated summaries," (b) "the variant without AI-generated summaries," or (c) "I don't have a preference." Immediately afterward, they were prompted to provide a brief justification: "Please justify your selection in 2–3 sentences or bullet points."

#### *Semi-structured Interview*

To contextualize the quantitative measures, we conducted a semi-structured interview (Adams, 2015) with each participant guided by three questions: (1) "If you think about both of your searches, what positive aspects stand out to you?", (2) "If you think about both of your searches, what negative aspects stand out to you?", (3) "Do you have any suggestions on how the AI-generated summaries could be made more useful for you?"

### Hypotheses

While prior evidence comes primarily from non-academic (web) search and single-document summarization, we extend this approach to multi-document summarization of academic search results. We expect comparable benefits, as information synthesis and relevance assessment are equally central in academic search, and synthesizing across multiple documents addresses a core challenge regarding the vast amount of scientific data.

**H1 (Subjective Workload)** Participants in the AISummary condition will report a lower subjective workload compared to participants in the NoSummary condition.
*Rationale:* The AI-generated summary aims to reduce mental effort by distilling information from multiple results. Prior work in non-academic contexts shows query-relevant summaries at the individual result level can shift effort from opening documents, enabling faster relevance judgments (Tombros & Sanderson, 1998). Query-biased summaries for individual search results associate with fewer queries, shorter times, and more relaxing, restful experiences (White et al., 2003). Non-conversational LLM-based tools show similar efficiency gains (Spatharioti et al., 2025). Google's "AI Overviews" did not reduce cognitive workload (Rolon-Merette et al., 2025), but our design and academic context (providing a domain-specific search engine) differ and might lead to different results.

**H2 (Perceived Usefulness)** Participants in the AISummary condition will rate the perceived usefulness of the search system higher than those in the NoSummary condition.
*Rationale:* The summary should enhance perceived system utility by scaffolding the search process. Users rated query-biased summaries for single results in non-academic search more useful than generic summaries (Alexander et al., 2000; White et al., 2003). Non-conversational LLM-based tools (Spatharioti et al., 2025) and Google's "AI Overviews" have been perceived as useful (Rolon-Merette et al., 2025).

**H3 (Overall Satisfaction)** Participants in the AISummary condition will report higher overall satisfaction with their search experience compared to those in the NoSummary condition.
*Rationale:* The summary aims to streamline search, potentially enhancing satisfaction. In non-academic web search, query-biased summaries of individual search results associate with higher satisfaction (White et al., 2003). Non-conversational LLM-based tools that accelerate task completion associate with enhanced perceived search experiences (Spatharioti et al., 2025).

**H4 (Decision-making Confidence)** Participants in the AISummary condition will report higher confidence in their decisions about which search results to explore further, compared to those in the NoSummary condition.
*Rationale:* The summary aims to help users assess result relevance, potentially increasing confidence in exploration decisions. Research in non-academic search shows that summaries' assessability, i.e., users' ability to tell whether results are useful without opening links, significantly influences which results users choose to explore (Mi & Jiang, 2019). This suggests that summaries could affect decision-making certainty about result selection.

### Participants

We recruited 30 participants (10 male, 20 female; ages 19–53, $M = 29.00$, $SD = 7.24$). 29 were recruited via the UK-based online crowdsourcing platform Prolific (https://www.prolific.com/), and one via LinkedIn. The participants were required to either be social scientists ($n = 10$) or social science students ($n = 20$), and to be fluent in English. Each participant was compensated £8.70 with an average completion time of approximately 43 minutes ($M = 42{:}57$ min; $SD = 07{:}02$ min), resulting in an hourly rate of roughly £12.14. The participants reported high familiarity with academic search engines overall ($M = 6.00$, $SD = 1.02$ on a 7-point Likert scale), and one participant reported to have been familiar with the search system before the study. We targeted 30 participants for this formative within-subjects study to identify directional patterns and elicit qualitative feedback, not to estimate small effects precisely.

### Results

In this section, we first present the results of our statistical analyses, including both confirmatory tests of our hypotheses and exploratory analyses of the NASA-TLX subscales. Subsequently, we show the results of our qualitative analyses from participants' post-task preference justifications and the semi-structured interviews.

### Statistical Analysis

We assessed normality of within-pair differences using the Shapiro–Wilk test. For normally distributed differences, we report paired-samples $t$-tests with Cohen's $d$ (Lakens, 2013). Otherwise, we report Wilcoxon signed-rank tests with Rosenthal's $r$ (Rosnow, 2003). We controlled family-wise error rate for the four confirmatory outcome measures (NASA–TLX total, perceived usefulness, satisfaction, confidence) using the Holm–Bonferroni procedure. Complete statistical results are presented in Table 1.

| Measure | NoSummary | AISummary | Median | Test | | | Effect | Diff.[c] |
|---|---|---|---|---|---|---|---|---|
| | *M (SD)* | *M (SD)* | (NS[a], AIS[b]) | Stat. | $p$ | $p_{adj}$ | Size[d] | 95% CI |

| *Confirmatory Outcomes* | | | | | | | | |
|---|---|---|---|---|---|---|---|---|
| NASA–TLX Total | 28.94 (20.75) | 23.69 (14.48) | 24.17, 21.67 | $t(29) = -1.598$ | .121 | .484 | $d = -0.292$ | [−11.97, 1.47] |
| Perceived Usefulness | 5.11 (1.51) | 5.39 (1.16) | 5.25, 5.58 | $W = 113.5$ | .465 | .787 | $r = +0.152$ | [−0.17, 0.17] |
| Overall Satisfaction | 4.90 (1.75) | 5.37 (1.52) | 5.00, 6.00 | $t(29) = 1.455$ | .156 | .484 | $d = +0.266$ | [−0.19, 1.12] |
| Decision Confidence | 5.13 (1.55) | 5.43 (1.61) | 5.00, 6.00 | $t(29) = 0.866$ | .393 | .787 | $d = +0.158$ | [−0.41, 1.01] |
| *Exploratory NASA–TLX Subscales* | | | | | | | | |
| Mental Demand | 41.67 (31.19) | 32.00 (25.35) | 42.50, 30.00 | $t(29) = -2.156$ | .040* | — | $d = -0.394$ | [−18.84, −0.49] |
| Frustration | 22.17 (31.34) | 12.17 (19.81) | 5.00, 2.50 | $W = 37.5$ | .068 | — | $r = -0.443$ | [−5.00, 0.00] |
| Performance | 42.33 (29.12) | 40.50 (28.81) | 42.50, 40.00 | $t(29) = -0.260$ | .796 | — | $d = -0.05$ | [−12.57, 16.23] |
| Effort | 39.83 (28.33) | 35.00 (26.26) | 35.00, 32.50 | $t(29) = -0.949$ | .351 | — | $d = -0.173$ | [−15.25, 5.58] |
| Temporal Demand | 17.00 (27.90) | 13.50 (17.87) | 5.00, 5.00 | $W = 79.0$ | .531 | — | $r = -0.144$ | [−2.50, 0.00] |
| Physical Demand | 10.67 (26.05) | 9.00 (20.27) | 0.00, 0.00 | $W = 44.0$ | .907 | — | $r = -0.03$ | [0.00, 0.00] |

* $p < .05$; [a]NS = *NoSummary*; [b]AIS = *AISummary*; [c]Difference calculated as *AISummary* - *NoSummary*.

[d]Effect size interpretations (Lakens, 2013; Rosnow, 2003): Cohen's $d$ (small = 0.2, medium = 0.5, large = 0.8); Rosenthal's $r$ (small = 0.1, medium = 0.3, large = 0.5).

**Table 1. Statistical Results for Confirmatory and Exploratory Analyses**

### *Confirmatory Analyses*

None of the four confirmatory measures reached statistical significance after Holm–Bonferroni adjustment (Table 1). Although all measures directionally favored *AISummary*, with lower workload, higher perceived usefulness, greater satisfaction, and increased confidence, the 95% confidence intervals (CIs) for all differences included zero, and observed effects were small. Thus, H1–H4 were not supported.

### *Exploratory NASA–TLX Subscales*

Following a hierarchical gatekeeping approach, we present NASA–TLX subscale analyses as exploratory, given the non-significant composite score. Mental demand showed a statistically significant reduction in the *AISummary* condition ($p = .040$, $d = -0.394$). Frustration did not reach conventional significance ($p = .068$, $r = -0.443$), with lower frustration in the *AISummary* condition. The remaining subscales showed no significant differences with negligible-to-small effects. These exploratory results suggest potential reductions in specific workload components with AI summaries.

### *AISummary Ratings & Interface Preference*

In general, participants evaluated the AI-generated summaries favorably on 7-point scales (with an "I don't know" option). The ratings were: *relevance assessability* ($M = 5.17$, $SD = 1.53$, $n = 26$), *appropriateness* in size and length ($M = 5.67$, $SD = 1.56$, $n = 28$), *understandability* ($M = 5.43$, $SD = 1.47$, $n = 27$), *perceived usefulness* ($M = 5.20$, $SD = 1.56$, $n = 27$), and *perceived reliability* ($M = 4.77$, $SD = 1.55$, $n = 25$). Regarding interface preference, 15 of the 30 participants favored the *AISummary* condition, eight preferred the *NoSummary* condition, and seven expressed no preference.

### *Behavioral Data*

| Condition | Topic | Result Clicks | Expansions | Bookmarks | Reformulations |
|---|---|---|---|---|---|
| | | $M$ ($SD$) | $M$ ($SD$) | $M$ ($SD$) | $M$ ($SD$) |
| *NoSummary* | Overall | 4.67 (3.70) | — | 5.87 (5.20) | 4.73 (4.13) |
| | Familiar | 5.81 (3.90) | — | 4.88 (3.34) | 5.38 (4.72) |
| | Predef. | 3.36 (3.08) | — | 7.00 (6.70) | 4.00 (3.35) |
| *AISummary* | Overall | 3.83 (3.84) | 0.97 (1.59) | 5.00 (3.93) | 4.27 (3.50) |
| | Familiar | 3.86 (4.29) | 1.14 (2.07) | 5.00 (4.71) | 5.71 (3.83) |
| | Predef. | 3.81 (3.54) | 0.81 (1.05) | 5.00 (3.27) | 3.00 (2.71) |

**Table 2. User Behavior by Condition and Topic Type**

As shown in Table 2, statistical comparisons of click counts did not yield significant differences between conditions, either overall or when stratified by topic type. In the *AISummary* condition, participants rarely expanded the AI summaries, suggesting that interaction was dominated by the default collapsed view in which the first four lines of the summaries were visible. Bookmarking behavior showed a similar pattern, with no significant differences between *NoSummary* and *AISummary* either overall or within topic types. Additionally, the number of query reformulations per task (i.e., modifications to a query while pursuing the same information need) did not differ between *NoSummary* and *AISummary* overall, nor within familiar or predefined topics (all $p > .05$). However, reformulations were significantly more frequent for familiar than unfamiliar topics (paired $t$-test $p = 0.011$).

### Qualitative Feedback

To analyze participants' responses to our open question to justify their interface preference selection as well as their interview answers regarding potential improvements of the AI-generated summary, we performed a thematic analysis (Braun & Clarke, 2006). Using inductive coding, two researchers (first and second authors of this paper) labeled all participant responses independently. It was permissible for a single response to be assigned multiple codes if it encompassed several distinct themes. Following the independent coding, the two researchers negotiated and resolved disagreements until consensus was reached for all responses, following best practices for negotiated agreement (McDonald et al., 2019).

#### Preference Justification

Among those favoring the interface featuring the AI-generated summary, two rationales dominated: *perceived speed, efficiency, and ease* ($n = 8$), and the value of a *high-level overview and starting point* ($n = 8$). Participants expressed reduced friction (e.g., Participant 6 (*P6*): *"It made the search so much easier and allowed me to find multiple journal articles relating to my topic easier"*) and the utility of a quick orientation (e.g., *P3*: *"The summary was nice and gave me a starting point at what to look at. But it didn't decrease my job at all."*). A third of this group highlighted *general search support and guidance* ($n = 5$), such as validating or refining keywords (*P13*: *"The AI overview helped me to determine if my keywords were the right words to find what I was looking for."*). Four participants noted that the summary supported *relevance judgments*: *P1* mentioned that it *"speeds up the process by allowing me to sift through and decide which articles are worthy to explore further." P11* wanted *deeper, more analytical functionality*: *"more elaborated information or analysis from AI not just the summary of top five most relevant papers."*

Participants who preferred the standard interface typically expressed AI reliability or visual design issues. *Trust, accuracy, and reliability* concerns were most prominent ($n = 5$): *P28* remarked, *"I wouldn't be sure to trust the AI overview without checking it myself so it feels redundant." Interface and visibility issues* also factored ($n = 3$), with *P14* stating, *"I didn't like the red color because it seemed like an error message."* Some simply preferred established habits (*P7*: *"Prefer to search for articles."*), while others framed *utility as context-dependent* (*P18*: *"AI-generated summaries of complex topics tend to be quite poor or superficial. […] It might be more useful for more unknown topics."*) or *not useful* at all (*P26*: *"[…] no need for AI to do it for me."*).

Those reporting no preference often described conditional value rather than blanket endorsement or rejection. The most common stance was *contextual utility* ($n = 3$): *P12* found the summary *"kinda useful"* for unfamiliar topics but negligible for familiar ones. *Interface and visibility issues* also mattered ($n = 2$), with some missing the feature entirely (*P8*: *"I didn't notice it"*). Two participants expressed *trust, accuracy, and reliability* reservations without it changing their preference (*P16*: *"I prefer to read the abstracts […] rather than trusting AI."*). Others perceived the summaries as *not useful* (*P24*: *"I do not find them particularly useful in general."*) or noted that while they appreciated the *general search support and guidance* provided, their search process was roughly comparable to what they could achieve unaided (*P27*: *"Overviews assured me which searching direct[ion] I should choose, but overall I think I will do as good without them."*).

Taken together, the justification patterns reveal a clear trade-off: users who valued efficiency, a focused starting point, and rapid relevance assessment leaned toward the summary, whereas users who prioritized verification, transparency, or who had strong existing preferences tended to see it as redundant or risky. Discoverability and visual appearance issues shaped perceptions at the margin.

#### Suggestions for Improving the AI-generated Summary

From participants' responses, we gathered 28 concrete suggestions (i.e., the total number of individual labels that emerged during the coding process; not every participant provided a suggestion), which cluster into presentation, transparency, interactivity, and user control. The most frequent request was a *more structured format* that supports scanning ($n = 8$), replacing a single dense paragraph with discrete units (e.g., bullet points, table structure, or thematic grouping): *P13* proposed *"five bullet points […] each one of them […] a separate link rather than it being all in one paragraph."* Close behind, six participants asked to *surface more metadata* such as authors, date/recency, geography, population, design (e.g., cross-sectional vs. longitudinal), key variables, and citation counts. *P28*, for example, wanted *"basic demographics of the sample […] the type of data […] and the key variables"* brought into the summary.

*Visual design fixes* ($n = 3$) targeted color issues and friction points such as hover interactions: *P18* suggested showing *"a little box just stating the name of the source"* while hovering over a reference. Calls for *more system transparency* ($n = 2$) centered on exposing confidence and ranking rationales. *P11* asked, *"[H]ow confident was the* *result? […] I'm also not expecting 95% […] but I'm also not expecting 5%."* Participants also desired *increased interactivity* to query the summary for specifics (*P17*: the ability to ask *"does it mention this?"* and receive a tailored *"resume"*), and *proactive recommendations* such as related studies and keyword hints (*P26*: *"suggestions that if you want to find something on this topic, you should type [xy]"*). Several asked for even *shorter, more concise summaries* (*P1*: *"maybe even more concise"*), while others emphasized user control and adherence to expectations: *make the feature optional* (*P18*), *group similar results and report counts within groups* (*P5*), and adopt a *more academic tone* (*P22*).

## DISCUSSION & IMPLICATIONS

This formative work extends prior research on query-biased summaries for single search results to SERP-level, multi-document AI summaries in academic search, where users seek source documents rather than direct answers.

First, in our manual metadata-grounded accuracy evaluation, GPT-4o mini received stronger labels than Llama 4 Scout. Qualitative analysis yielded a six-category exploratory error taxonomy including (i) inaccurate or unsupported content descriptions, (ii) unsubstantiated guesses, (iii) omitted search results, (iv) incorrect identifiers, (v) formatting and language issues, and (vi) inappropriate grouping of results. Model choice thus seems to function as an interaction-design variable, with errors compromising identifier fidelity or result coverage posing particular threats in scholarly contexts where citation integrity is essential. To address these failure modes, we propose five safeguards: (i) implement pre-render coverage checks that flag missing items, (ii) validate identifiers via pattern-matching or database lookups, (iii) use template-based formatting and employ few-shot prompting to reduce formatting errors, (iv) tighten prompt constraints against speculation and illogical grouping, and (v) add a lightweight verifier LLM to audit identifiers and search result coverage before display.

Second, the user study provides qualified evidence of benefit rather than blanket endorsement. Confirmatory outcomes consistently trended in favor of the AI-summary interface but remained non-significant with small effects, while exploratory analyses suggested lower mental demand. Together with favorable ratings of understandability, usefulness, and relevance assessability, as well as qualitative accounts of rapid orientation and relevance assessment support, these patterns suggest that SERP-level summaries may help users orient more quickly without replacing source inspection. Interpreted through Information Foraging Theory, we propose that the summary can act as a "scent concentrator": it compresses cues from multiple top-ranked results into a compact, source-linked representation that may support early-stage triage by reducing the cognitive cost of entering the information patch. Rare expansion of the summary suggests that the collapsed view may have provided enough SERP-level scent for coarse orientation, while abstracts and metadata remained available for deeper judgment of individual results.

At the same time, preferences were mixed, underscoring that the AI summaries are best understood as a context- and user-dependent aid rather than a universal improvement. Their value may lie in reducing some early-stage triage effort for some users and tasks, particularly when they are designed for scanability, transparency, responsiveness, and user control.

## LIMITATIONS

This study has several limitations. First, the accuracy evaluation covered only ten queries, so the error taxonomy is exploratory and should be validated on broader query sets, domains, and models. Second, the sample size offered limited power for detecting small effects. Most quantitative differences therefore remained inconclusive and should be interpreted as hypothesis-generating rather than confirmatory. Third, most participants were recruited via Prolific, sessions were remote and time-limited, and participant background, AI-tool use, and topic familiarity were not modeled. Fourth, the summary always covered the top five results, so we cannot assess alternative coverage thresholds or dynamic cutoffs. We also lacked expert relevance judgments and therefore cannot assess retrieval effectiveness directly. Finally, the study focused on the social sciences, limiting generalizability.

## CONCLUSION

In a formative mixed-methods design study, we examined whether a SERP-level AI summary of the five top-ranked results can support information triage during academic search in the social sciences. In a manual metadata-grounded accuracy evaluation, GPT-4o mini received stronger labels than Llama 4 Scout and revealed practically important risks for scholarly use, motivating safeguards such as coverage checks and identifier validation. In the user study, outcomes consistently trended in favor of the AI-summary interface, with exploratory evidence of lower mental demand, while behavioral patterns suggested slightly fewer clicks and reformulations. Interpreted through Information Foraging Theory, these findings suggest that compact, source-linked summaries may concentrate SERP-level information scent and support early-stage triage without replacing source inspection. Overall, overview-level summarization appears to be a context- and user-dependent aid rather than a universal improvement. Future work should test this account with larger and longitudinal studies, examine applicability beyond the social sciences, vary summary length and coverage, and include direct performance measures such as relevance of bookmarked items and time to useful sources alongside attention-sensitive methods such as eye tracking.

## GENERATIVE AI USE

We employed ChatGPT, Claude, and Gemini for the following purposes: improving the wording of paragraphs, helping with the creation of tables, and helping with the creation of data analysis scripts. We evaluated the output by carefully checking for factual correctness and logical consistency. The authors assume all responsibility for the content of this submission.

## AUTHOR ATTRIBUTION

First Author: conceptualization, methodology, investigation, formal analysis, writing – original draft; Second Author: investigation, formal analysis, conceptualization, writing – review and editing; Third Author: project administration, conceptualization, writing – review and editing; Fourth Author: project administration, investigation, writing – review and editing; Fifth Author: project administration, funding acquisition, supervision, conceptualization, methodology, formal analysis, writing – review and editing; Sixth Author: project administration, funding acquisition, conceptualization, methodology, software, formal analysis, writing – review and editing.

## ACKNOWLEDGEMENTS

This work is funded by the German Research Foundation (DFG) project "VACOS 2" project (no. 388815326). Furthermore, this work is supported by the EU-funded "OMINO" project (DOI: 10.3030/101086321). We thank our study participants and the reviewers for their valuable feedback.

## REFERENCES

Abdul Salam, M., Gamal, M., Hamed, H. F. A., & Sweidan, S. (2025). Abstractive text summarization using deep learning models: A survey. *International Journal of Data Science and Analytics*, *20*(5), 4209–4237. https://doi.org/10.1007/s41060-025-00743-w

Adams, W. C. (2015). Conducting semi-structured interviews. In *Handbook of practical program evaluation* (pp. 492–505). John Wiley & Sons, Ltd. https://doi.org/10.1002/9781119171386.ch19

Al Lawati, S. F. D. (2025). I am not a caveman: An eye-tracking study of how users are influenced to search in the era of GenAI. *Companion Proceedings of the ACM on Web Conference 2025*, 681–684. https://doi.org/10.1145/3701716.3715284

Alanzi, E., & Alballaa, S. (2023). Query-focused multi-document summarization survey. *International Journal of Advanced Computer Science and Applications*, *14*(6). https://doi.org/10.14569/IJACSA.2023.0140688

Alexander, N., Brown, C., Jose, J. M., Ruthven, I., & Tombros, A. (2000). Question answering, relevance feedback and summarisation: TREC-9 interactive track report. In E. M. Voorhees & D. K. Harman (Eds.), *Proceedings of the ninth text REtrieval conference, TREC 2000, gaithersburg, maryland, USA, november 13-16, 2000* (Vols. 500–249). National Institute of Standards; Technology (NIST). http://trec.nist.gov/pubs/trec9/papers/glasgow_proceedings.pdf

Bawden, D., & Robinson, L. (2009). The dark side of information: Overload, anxiety and other paradoxes and pathologies. *Journal of Information Science*, *35*(2), 180–191. https://doi.org/10.1177/0165551508095781

Belabbes, M. A., Ruthven, I., Moshfeghi, Y., & Rasmussen Pennington, D. (2022). Information overload: A concept analysis. *Journal of Documentation*, *79*(1), 144–159. https://doi.org/10.1108/JD-06-2021-0118

Braun, V., & Clarke, V. (2006). Using thematic analysis in psychology. *Qualitative Research in Psychology*, *3*(2), 77–101. https://doi.org/10.1191/1478088706qp063oa

Cachola, I., Lo, K., Cohan, A., & Weld, D. (2020). TLDR: Extreme summarization of scientific documents. In T. Cohn, Y. He, & Y. Liu (Eds.), *Findings of the association for computational linguistics: EMNLP 2020: Vols. EMNLP 2020* (pp. 4766–4777). Association for Computational Linguistics. https://doi.org/10.18653/v1/2020.findings-emnlp.428

Chen, J., Lin, H., Han, X., & Sun, L. (2024). Benchmarking large language models in retrieval-augmented generation. In M. J. Wooldridge, J. G. Dy, & S. Natarajan (Eds.), *Proceedings of the thirty-eighth AAAI conference on artificial intelligence and thirty-sixth conference on innovative applications of artificial intelligence and fourteenth symposium on educational advances in artificial intelligence* (pp. 17754–17762). AAAI Press. https://doi.org/10.1609/aaai.v38i16.29728

Chu, B., Li, M., Frihat, S., Gu, C., Lodde, G., Livingstone, E., & Fuhr, N. (2025). *TracSum: A new benchmark for aspect-based summarization with sentence-level traceability in medical domain*. https://arxiv.org/abs/2508.13798

Davis, F. D. (1989). Perceived usefulness, perceived ease of use, and user acceptance of information technology. *MIS Quarterly*, *13*(3), 319–340. https://doi.org/10.2307/249008

El-Kassas, W. S., Salama, C. R., Rafea, A. A., & Mohamed, H. K. (2021). Automatic text summarization: A comprehensive survey. *Expert Systems with Applications*, *165*, 113679. https://doi.org/10.1016/j.eswa.2020.113679

Extance, A. (2018). How AI technology can tame the scientific literature. *Nature*, *561*(7722), 273–274.

Godbole, A., George, J. G., & Shandilya, S. (2025). Leveraging long-context large language models for multi-document understanding and summarization in enterprise applications. In V. B. Gupta, S. K. Shandilya, F. Ortiz-Rodríguez, & J. L. Martinez-Rodriguez (Eds.), *Business intelligence, computational mathematics, and data analytics* (pp. 208–224). Springer Nature Switzerland.

Grainger, T., Turnbull, D., & Irwin, M. (2025). *AI-powered search*. Simon; Schuster.

Gulati, S., Sousa, S., & Lamas, D. (2019). Design, development and evaluation of a human-computer trust scale. *Behav. Inf. Technol.*, *38*(10), 1004–1015. https://doi.org/10.1080/0144929X.2019.1656779

Gusenbauer, M. (2019). Google scholar to overshadow them all? Comparing the sizes of 12 academic search engines and bibliographic databases. *Scientometrics*, *118*(1), 177–214. https://doi.org/10.1007/s11192-018-2958-5

Hart, S. G., & Staveland, L. E. (1988). Development of NASA-TLX (task load index): Results of empirical and theoretical research. In P. A. Hancock & N. Meshkati (Eds.), *Human mental workload* (Vol. 52, pp. 139–183). North-Holland. https://doi.org/10.1016/S0166-4115(08)62386-9

Hołyst, J. A., Mayr, P., Thelwall, M., Frommholz, I., Havlin, S., Sela, A., Kenett, Y. N., Helic, D., Rehar, A., Maček, S. R., Kazienko, P., Kajdanowicz, T., Biecek, P., Szymanski, B. K., & Sienkiewicz, J. (2024). Protect our environment from information overload. *Nature Human Behaviour*, *8*(3), 402–403. https://doi.org/10.1038/s41562-024-01833-8

Ibrahim Altmami, N., & El Bachir Menai, M. (2022). Automatic summarization of scientific articles: A survey. *Journal of King Saud University - Computer and Information Sciences*, *34*(4), 1011–1028. https://doi.org/10.1016/j.jksuci.2020.04.020

Kacprzak, E., Koesten, L., Ibáñez, L.-D., Blount, T., Tennison, J., & Simperl, E. (2019). Characterising dataset search—an analysis of search logs and data requests. *Journal of Web Semantics*, *55*, 37–55. https://doi.org/10.1016/j.websem.2018.11.003

Kaiser, C., Kaiser, J., Schallner, R., & Schneider, S. (2025). A new era of online search? A large-scale study of user behavior and personal preferences during practical search tasks with generative AI versus traditional search engines. *Proceedings of the Extended Abstracts of the CHI Conference on Human Factors in Computing Systems*. https://doi.org/10.1145/3706599.3720123

Kuhlthau, C. C. (1991). Inside the search process: Information seeking from the user's perspective. *Journal of the American Society for Information Science*, *42*(5), 361–371. https://doi.org/10.1002/(SICI)1097-4571(199106)42:5<361::AID-ASI6>3.0.CO;2-\#

Lakens, D. (2013). Calculating and reporting effect sizes to facilitate cumulative science: A practical primer for t-tests and ANOVAs. *Frontiers in Psychology*, *4*, 863. https://doi.org/10.3389/fpsyg.2013.00863

Lee, H.-P. (Hank)., Sarkar, A., Tankelevitch, L., Drosos, I., Rintel, S., Banks, R., & Wilson, N. (2025). The impact of generative AI on critical thinking: Self-reported reductions in cognitive effort and confidence effects from a survey of knowledge workers. *Proceedings of the 2025 CHI Conference on Human Factors in Computing Systems, CHI 2025, YokohamaJapan, 26 April 2025- 1 May 2025*, 1121:1–1121:22. https://doi.org/10.1145/3706598.3713778

Li, J., Tang, Z., Liu, X., Spirtes, P., Zhang, K., Leqi, L., & Liu, Y. (2024). Steering LLMs towards unbiased responses: A causality-guided debiasing framework. *arXiv Preprint arXiv:2403.08743*. https://doi.org/10.48550/arXiv.2403.08743

Madsen, M., & Gregor, S. (2000). Measuring human-computer trust. *11th Australasian Conference on Information Systems*, *53*, 6–8.

Mao, Y., Zhong, M., & Han, J. (2022). CiteSum: Citation text-guided scientific extreme summarization and domain adaptation with limited supervision. In Y. Goldberg, Z. Kozareva, & Y. Zhang (Eds.), *Proceedings of the 2022 conference on empirical methods in natural language processing* (pp. 10922–10935). Association for Computational Linguistics. https://doi.org/10.18653/v1/2022.emnlp-main.750

Marshall, C. C., & Shipman, F. M. (1997). Spatial hypertext and the practice of information triage. *Proceedings of the Eighth ACM Conference on Hypertext*, 124–133. https://doi.org/10.1145/267437.267451

Maxwell, D., Azzopardi, L., & Moshfeghi, Y. (2017). A study of snippet length and informativeness: Behaviour, performance and user experience. *Proceedings of the 40th International ACM SIGIR Conference on Research and Development in Information Retrieval*, 135–144. https://doi.org/10.1145/3077136.3080824

Mayerhofer, K., Capra, R., & Elsweiler, D. (2025). Blending queries and conversations: Understanding trust, verification, and system choice in search and chat interactions. *Proceedings of the 2025 ACM SIGIR Conference on Human Information Interaction and Retrieval*, 168–178. https://doi.org/10.1145/3698204.3716454

McDonald, N., Schoenebeck, S., & Forte, A. (2019). Reliability and inter-rater reliability in qualitative research: Norms and guidelines for CSCW and HCI practice. *Proc. ACM Hum.-Comput. Interact.*, *3*(CSCW), 72:1–72:23. https://doi.org/10.1145/3359174

Meng, R., Thaker, K., Zhang, L., Dong, Y., Yuan, X., Wang, T., & He, D. (2021). Bringing structure into summaries: A faceted summarization dataset for long scientific documents. In C. Zong, F. Xia, W. Li, & R. Navigli (Eds.), *Proceedings of the 59th annual meeting of the association for computational linguistics and the 11th international joint conference on natural language processing (volume 2: Short papers)* (pp. 1080–1089). Association for Computational Linguistics. https://doi.org/10.18653/v1/2021.acl-short.137

Mi, S., & Jiang, J. (2019). Understanding the interpretability of search result summaries. *Proceedings of the 42nd International ACM SIGIR Conference on Research and Development in Information Retrieval*, 989–992. https://doi.org/10.1145/3331184.3331306

Narayanan Venkit, P., Laban, P., Zhou, Y., Mao, Y., & Wu, C.-S. (2025). Search engines in the AI era: A qualitative understanding to the false promise of factual and verifiable source-cited responses in LLM-based search. *Proceedings of the 2025 ACM Conference on Fairness, Accountability, and Transparency*, 1325–1340. https://doi.org/10.1145/3715275.3732089

Ortega, J. L. (2014). *Academic search engines*. Chandos Publishing (Oxford). https://doi.org/10.1016/C2013-0-23226-8

Papenmeier, A., Krämer, T., Friedrich, T., Hienert, D., & Kern, D. (2021). Genuine information needs of social scientists looking for data. *Proceedings of the Association for Information Science and Technology*, *58*(1), 292–302. https://doi.org/10.1002/pra2.457

Roetzel, P. G. (2019). Information overload in the information age: A review of the literature from business administration, business psychology, and related disciplines with a bibliometric approach and framework development. *Business Research*, *12*(2), 479–522. https://doi.org/10.1007/s40685-018-0069-z

Rolon-Merette, T., Guo, E. Y., Poivré, F., Benbousta, M., Amoros, S., Karran, A. J., Coursaris, C. K., Senecal, S., & Leger, P.-M. (2025). Generative AI Summaries Do Not Increase User Engagement: A Study of Google AI Overviews. *Proceedings of the SIGHCI 2024 Workshop on HCI Research in MIS*. https://aisel.aisnet.org/sighci2024/29

Rosnow, R. L. (2003). Effect sizes for experimenting psychologists. *Canadian Journal of Experimental Psychology*, *57*(3), 221–237.

Schütze, H., Manning, C. D., & Raghavan, P. (2008). *Introduction to information retrieval* (Vol. 39). Cambridge University Press Cambridge.

Spatharioti, S. E., Rothschild, D., Goldstein, D. G., & Hofman, J. M. (2025). Effects of LLM-based search on decision making: Speed, accuracy, and overreliance. *Proceedings of the 2025 CHI Conference on Human Factors in Computing Systems*. https://doi.org/10.1145/3706598.3714082

Sullivan, V., & Weger, K. (2025). Transparency and explainability in AI-assisted decision making: Effects on trust, perceived reliability, confidence, and ease of understanding. *Proc. Hum. Factors Ergon. Soc. Annu. Meet.*, *10711813251369473*. https://doi.org/10.1177/10711813251369473

Sweeney, S., & Crestani, F. (2006). Effective search results summary size and device screen size: Is there a relationship? *Information Processing & Management*, *42*(4), 1056–1074. https://doi.org/10.1016/j.ipm.2005.06.007

Tenopir, C., King, D. W., Boyce, P., Grayson, M., Zhang, Y., & Ebuen, M. (2003). Patterns of journal use by scientists through three evolutionary phases. *DLib Mag.*, *9*(5). https://doi.org/10.1045/may2003-king

Tombros, A., & Sanderson, M. (1998). Advantages of query biased summaries in information retrieval. *Proceedings of the 21st Annual International ACM SIGIR Conference on Research and Development in Information Retrieval*, 2–10. https://doi.org/10.1145/290941.290947

White, R. W., Jose, J. M., & Ruthven, I. (2003). A task-oriented study on the influencing effects of query-biased summarisation in web searching. *Information Processing & Management*, *39*(5), 707–733. https://doi.org/10.1016/S0306-4573(02)00033-X

Withorn, T. (2025). Google AI overviews are here to stay: A call to teach AI literacy. *College & Research Libraries News*, *86*(5), 214. https://doi.org/10.5860/crln.86.5.214

Yang, Y., Urgo, K., Arguello, J., & Capra, R. (2025). Search+chat: Integrating search and GenAI to support users with learning-oriented search tasks. *Proceedings of the 2025 ACM SIGIR Conference on Human Information Interaction and Retrieval*, 57–70. https://doi.org/10.1145/3698204.3716446

Zaman, F., Kamiran, F., Shardlow, M., Hassan, S.-U., Karim, A., & Aljohani, N. R. (2024). SATS: Simplification aware text summarization of scientific documents. *Frontiers Artif. Intell.*, *7*, 1375419. https://doi.org/10.3389/FRAI.2024.1375419

Zhang, H., Yu, P. S., & Zhang, J. (2025). A systematic survey of text summarization: From statistical methods to large language models. *ACM Comput. Surv.*, *57*(11). https://doi.org/10.1145/3731445

Zhang, Y., Jin, H., Meng, D., Wang, J., & Tan, J. (2025). *A comprehensive survey on process-oriented automatic text summarization with exploration of LLM-based methods*. https://arxiv.org/abs/2403.02901